\def\cmmoinsdeux{\mbox{ cm}^{-2}}

\def\microns{\mbox{ } \mu \mbox{m}}

\def\mags{\mbox{ magnitudes}}

\def\adeg{^{\circ}}

\def\amin{^\prime}

\def\nh{N_{\rm H}}

\def\ltsima{\; \buildrel < \over \sim \;}
\def\simlt{\lower.5ex\hbox{\ltsima}}            % < over MMM
\def\gtsima{\; \buildrel > \over \sim \;}
\def\simgt{\lower.5ex\hbox{\gtsima}}            % > over MMM

\documentclass[
    ,final            % use final for the camera ready runs
  ]
  {aipproc}

\layoutstyle{8x11single}

\begin{document}

\title{How to reveal the mysteries of the most obscured high-energy
sources of our Galaxy, discovered by {\it INTEGRAL}?}


\author{Sylvain Chaty}{
  address={AIM - Astrophysique Interactions Multi-échelles 
(UMR 7158 CEA/CNRS/Université Paris 7 Denis Diderot)
CEA Saclay, DSM/DAPNIA/Service d'Astrophysique, Bât. 709, L'Orme des Merisiers
FR-91 191 Gif-sur-Yvette Cedex, France,
%Tel:+33 1 69 08 43 85 / Fax:+33 1 69 08 65 77 / 
{\em chaty@cea.fr}
}
}
%\author{<author2>}{
%  address={<common address for author2 and author3>}
%}
%\author{<author3>}{
%  address={<common address for author2 and author3>}
%  ,altaddress={<author1 address>} % additional visiting address
%}

\begin{abstract}
A new type of high-energy binary systems has been revealed by the {\it
INTEGRAL} satellite.  These sources are in the course of being
unveiled by means of multi-wavelength optical, near- and mid-infrared
observations. Among these sources, two distinct classes are appearing:
the first one is constituted of intrinsically obscured high-energy
sources, of which IGR~J16318-4848 seems to be the most extreme example. The
second one is populated by the so-called supergiant fast X-ray
transients, with IGR~J17544-2619 being the archetype. We report here
on multi-wavelength optical to mid-infrared observations of these
systems. We show that in the case of the obscured sources our
observations suggest the presence of absorbing material (dust and/or
cold gas) enshrouding the whole binary system.  We then discuss the
nature of these two different types of systems.
\end{abstract}

\maketitle

%%%%%%%%%%%%%%%%%%%%%%%%%%%%%%%%%%%%%%%%%%%%
%% MAINMATTER
%%%%%%%%%%%%%%%%%%%%%%%%%%%%%%%%%%%%%%%%%%%%

\section{Introduction}

The {\it INTEGRAL} observatory has performed a detailed survey of the
galactic plane and the ISGRI detector on the IBIS imager has
discovered many new high energy sources, most of all reported in
\cite{bird:2006} (and
http://isdc.unige.ch/$\sim$rodrigue/html/igrsources\-.html).  The most
important result of {\it INTEGRAL} to date is the discovery of many
new high energy sources --concentrated in the Galactic plane, and some
in the Norma arm (see e.g. \cite{chaty:2005a})--,
%\citeyear{chaty:2005a})--,
%and \citeauthor{tomsick:2004a} \citeyear{tomsick:2004a})--, 
exhibiting common characteristics which previously had rarely been
seen.  Most of them are high mass X-ray binaries (HMXBs) hosting a
neutron star orbiting around an O/B companion, in some cases a
supergiant star. They divide into two classes: some of the new
sources are very obscured, and exhibiting a huge intrinsic and local
extinction, 
%--the most extreme example is the highly absorbed source IGR~J16318-4848
%\citep{filliatre:2004}--, 
and the others are HMXBs hosting a
supergiant star and exhibiting fast and transient outbursts: an unusual
characteristic among HMXBs: they are therefore called Supergiant Fast
X-ray Transients (SFXTs, \cite{negueruela:2006a}).
%, with IGR~J17544-2619 being their archetype.
%\citeyear{negueruela:2006a})--.
%
High-energy observations are not sufficient to reveal the nature of
the newly discovered sources, since the {\it INTEGRAL} localisation
($\sim 2\amin$) is not accurate enough to unambiguously pinpoint the
source at other wavelengths. Once X-ray satellites such as {\it
XMM-Newton}, {\it Chandra} or {\it Swift} provide an arcsecond
position, the hunt for the optical counterpart of the source is open.
However, the high level of absorption towards the galactic plane makes
the near-infrared (NIR) domain more efficient to identify these
sources.  We first report on multi-wavelength observations of two
sources belonging to each class described above, then give general
results on {\it INTEGRAL} sources, before
discussing them and concluding.
%in Section \ref{conclusions}.

\section{Observations and results} \label{observations}

The multiwavelength observations described here were performed
at the European Southern Observatory (ESO), using
Target of Opportunity (ToO) and Visitor modes, in 3 domains: optical
($400-800 \microns$) with the EMMI instrument on the
3.5m New Technology Telescope (NTT) at La Silla, NIR 
($1-2.5 \microns$) with the SOFI instrument on the NTT, and MIR
($5-20 \microns$) with the VISIR instrument on Melipal,
the 8m Unit Telescope 3 (UT3) of the Very Large Telescope (VLT) at
Paranal (Chile).  These observations include photometry and spectroscopy on 20
{\it INTEGRAL} sources in order to identify their counterparts, the
nature of the companion star, derive the distance, and finally
characterise the presence and temperature of their circumstellar
medium.

     \subsection{IGR~J16318-4848: extreme among the obscured high-energy
sources}

IGR~J16318-4848 was the first source to be discovered by IBIS/ISGRI
on {\it INTEGRAL} on 29 January 2003
\cite{courvoisier:2003}. {\it XMM-Newton} 
observations showed 
%that it was exhibiting 
a strong absorption
of $\nh \sim 2 \times 10^{24} \cmmoinsdeux$ \cite{matt:2003}.  The
accurate localisation by {\it XMM-Newton} allowed \cite{filliatre:2004} to rapidly
trigger ToO photometric and spectroscopic observations in optical and
NIR, leading to the discovery of the optical counterpart and to the
confirmation of the NIR one found by \cite{walter:2003}.  The 
extremely bright NIR source
%(J\,$= 10.33\pm 0.14$; H\,$=8.33\pm 0.10$ and
(Ks\,$=7.20 \mags$) %, and 
exhibits an unusually strong intrinsic
absorption %in the optical 
of $A_v = 17.4 \mags$, much stronger than
the absorption along the line of sight of $A_v = 11.4
\mags$, but still 100 times lower than the absorption in X-rays!  This
led \cite{filliatre:2004} to suggest that the material absorbing in
the X-rays was concentrated around the compact object, while the
material absorbing in the optical/NIR was enshrouding the
whole system.  The NIR spectroscopy %in the $0.95-2.5 \microns$ domain
revealed an unusual spectrum, with many strong emission lines,
originating from a highly complex and stratified circumstellar
environment, of various densities and temperatures, suggesting the
presence of an envelope and strong stellar outflow, responsible for the
absorption. Only luminous early-type stars such as supergiant sgB[e] 
show such extreme environments, and
\cite{filliatre:2004} concluded that IGR~J16318-4848 was an unusual
HMXB.
Combining these optical and NIR data with MIR observations, and
fitting these observations with a model of a sgB[e] companion star,
allowed \cite{rahoui:2007} to show that IGR~J16318-4848 exhibits a MIR
excess (see Figure \ref{figure:igrj16318-igrj17544}), that they
interpret as being due to the strong stellar outflow emanating from
the sgB[e] companion star.  They found that the companion star had a
temperature of T\,$=23500$\,K, radius R$_{star} = 20.4 R_{\odot}$, and
an extra component of temperature T $=900$\,K, radius R\,$= 12 R_{star}$
and A$_v = 17.6
\mags$. The extension of this extra component
seems to suggest that it enshrouds the whole binary system, as would
do a cocoon of gas/dust.
In summary, IGR~J16318-4848 is an HMXB system, located at a distance
between 1 to 6 kpc, hosting a neutron star (probably) and a sgB[e] star,
%It is therefore the second HMXB with a sgB[e] star, after CI Cam.  
%the most striking fact being that 
%i) the compact object seems to be surrounded by absorbing material and ii)
and the whole binary system seems to be surrounded by a dense and
absorbing circumstellar material envelope or cocoon, made of cold gas
and/or dust. This source exhibits so extreme characteristics that
it might not be fully representative of the other obscured sources. 

     \subsection{IGR~J17544-2619: archetype of the Supergiant Fast
X-ray Transients}

\begin{figure}
  \includegraphics[height=.3\textheight,angle=-90]{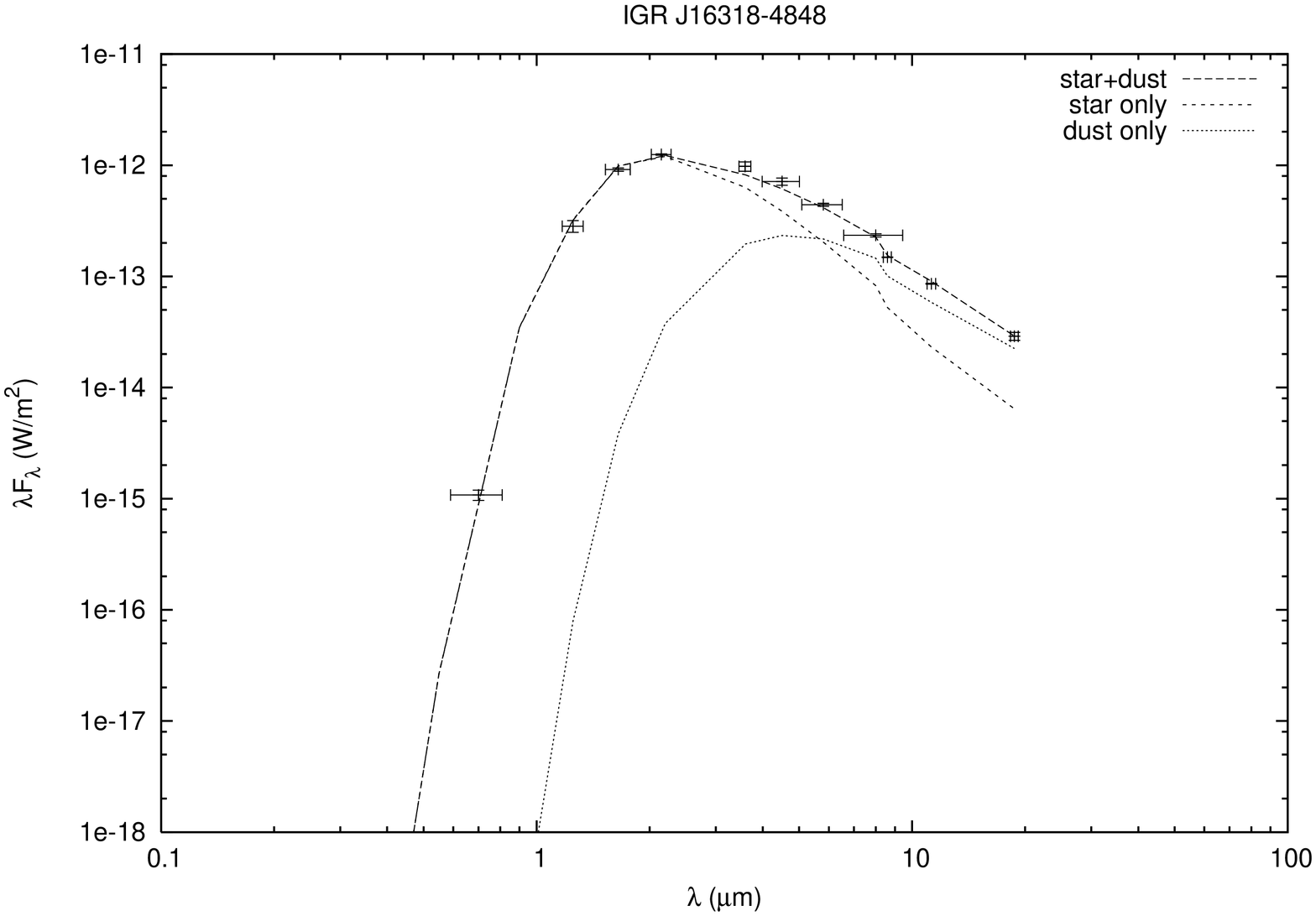}
  \includegraphics[height=.3\textheight,angle=-90]{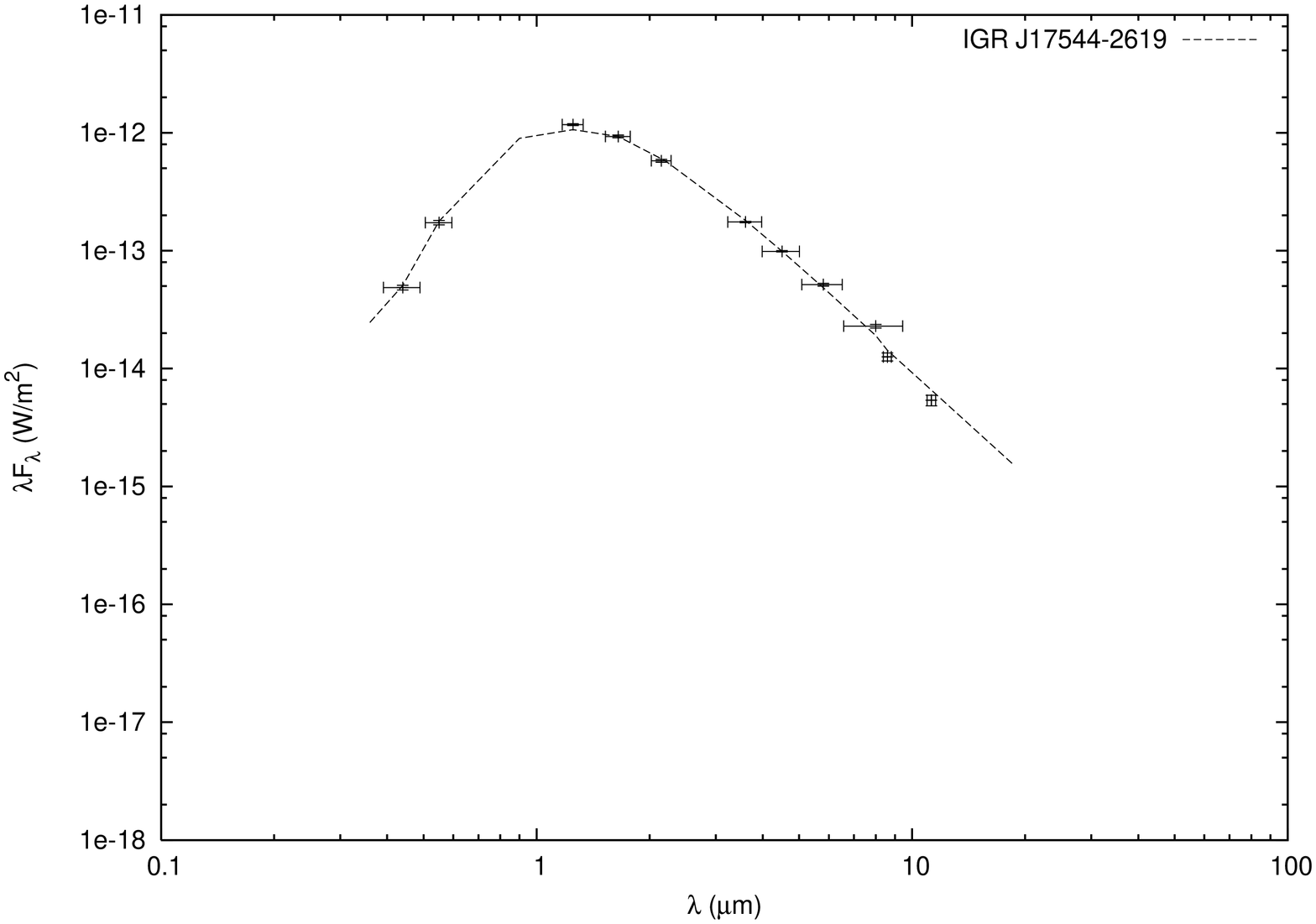}
  \caption{\label{figure:igrj16318-igrj17544} Optical to MIR SEDs of
  IGR~J16318-4848 (left) and 
IGR~J17544-2619 (right), including data from
  ESO/NTT, VISIR on VLT/UT3 and {\it Spitzer} \cite{rahoui:2007}. 
IGR~J16318-4848 exhibits a MIR excess, interpreted by \cite{rahoui:2007}
  as the signature of a strong stellar outflow coming from the sgB[e]
  companion star \cite{filliatre:2004}.  On the other hand, 
IGR~J17544-2619 is well fitted with only a stellar component
  corresponding to the companion star spectral type (O9Ib)
  \cite{pellizza:2006}.}
\end{figure}

The Supergiant Fast X-ray Transients (SFXTs) constitute a new class of
sources identified among the recently discovered {\it INTEGRAL}
sources, whose common characteristics are: they exhibit rapid
outbursts lasting only hours, a faint quiescent emission, their high
energy spectra require a BH or NS accretor, and they host O/B
supergiant companion stars.  Among these sources, IGR~J17544-2619, a
bright recurrent transient X-ray source discovered by {\it INTEGRAL}
on 17 September 2003 \citep{sunyaev:2003b}, seems to be the
archetype. Observations with {\it XMM-Newton} have shown that it
exhibits a very hard X-ray spectrum, and a faint intrinsic
absorption ($10^{22} \cmmoinsdeux$)
\cite{gonzalez-riestra:2004}.  Its bursts last for hours, in-between
bursts it exhibits long quiescence periods, and a long
outburst period of 165 days \citep{negueruela:2006a}. The nature of
the compact object is probably a neutron star \cite{intzand:2005}. 
\cite{pellizza:2006} managed to get optical/NIR ToO 
observations only one day after the discovery of this source. They
identified a likely counterpart inside the {\it XMM-Newton} error
circle, confirmed by {\it Chandra} accurate localization.
Spectroscopy showed that the companion star was a blue supergiant of
spectral type O9Ib, with a mass of $25-28 M_{\odot}$ and temperature
of T $\sim 31000$ K: the system is therefore an HMXB \cite{pellizza:2006}.
%, at a distance of 3-4 kpc .
\cite{rahoui:2007} combined optical, NIR and MIR observations
and showed that they could accurately fit the observations with a
model of an O9Ib star: temperature T~$=30500$~K and radius R$_{star} =
21.9 R_{\odot}$.  The absorption they derived was
A$_v = 5.9 \mags$ and the distance D~$=3.9$~kpc. 
The source does not exhibit any MIR excess
(see Figure \ref{figure:igrj16318-igrj17544}) \cite{rahoui:2007}.
In summary, IGR~J17544-2619 is a HMXB at a distance of $\sim$4~kpc,
constituted of an O9Ib supergiant, 
with a mild stellar wind and a compact object which is
probably a neutron star, without any absorbing material.

%---------------------------------------------------------------------
%\includegraphics[height=.3\textheight]{/Users/chaty/Presentations/Congres/Integral/Moscou/Proceedings/HAL/igr17544.ps}
%---------------------------------------------------------------------

     \subsection{General results on {\it INTEGRAL} sources and discussion} \label{IGRs}

In order to better characterize this population, 
\cite{chaty:2007b} and \cite{rahoui:2007} have studied a sample
of 20 {\it INTEGRAL} sources belonging to both classes described above.
The optical/NIR study allowed \cite{chaty:2007b} to identify or confirm the
identification of the counterpart, and to show that most of these
systems are HMXBs, containing massive and luminous early-type
companion stars.  By fitting the spectral energy
distributions of these sources from optical to MIR,
\cite{rahoui:2007} showed that i. most of them exhibit an intrinsic
absorption and ii. three of them exhibit
a MIR excess, that they suggest to be due to the presence of a cocoon of
dust and/or cold gas enshrouding the whole binary system
(see also \cite{chaty:2006c}).
%
%\section{Conclusions} \label{conclusions}
%
These results confirm the existence in the Galaxy of a dominant
population of a previously rare class of high-energy binary systems,
constituted of supergiant HMXBs with
high intrinsic absorption for some of them \cite{chaty:2007b}
\cite{rahoui:2007}.
%has been recently revealed by the {\it INTEGRAL} high-energy observatory .  
Fundamental differences exist between obscured sources and SFXTs, and
one possibility to explain those is to invoke a different geometry of
the binary systems, or a different extension of a wind/cocoon enshrouding
either the companion star or the whole system \cite{chaty:2006c}.  It
is now clear that a careful study of this new population will
provide a better understanding of the formation and evolution of 
%high energy binary systems and 
such short-living HMXBs of our Galaxy, and
will allow in the future stellar population models to better take
these systems into account, to assess a realistic number of
high-energy binary systems in our Galaxy.  Our final word is that the
GLAST satellite will certainly discover such new and unexpected
objects, and that, as for these obscured high-energy sources, only a
multiwavelength study will allow to reveal their nature.

%%%%%%%%%%%%%%%%%%%%%%%%%%%%%%%%%%%%%%%%%%%%%%%%
%% BACKMATTER
%%%%%%%%%%%%%%%%%%%%%%%%%%%%%%%%%%%%%%%%%%%%%%%%

\begin{theacknowledgments}
SC would like to thank the organisers for the opportunity to report on
these exciting results on newly discovered {\it INTEGRAL} sources and
for organizing a press conference on this subject.  SC is grateful to
Juan-Antonio Zurita Heras for useful comments on the manuscript,
and to Farid Rahoui for making the SEDs of Figure 1.
Based on observations collected at the European Southern Observatory,
Chile (proposals ESO N$\adeg$ 070.D-0340, 071.D-0073, 073.D-0339,
075.D-0773 and 077.D-0721).
%, and also for a very nice
%organisation of this workshop, fruitful to arise scientific
%discussions and new ideas.
\end{theacknowledgments}



\bibliographystyle{aipproc}   % if natbib is available

%\bibliography{/Users/chaty/Library/Texmf/Science/science}

\IfFileExists{\jobname.bbl}{}
 {\typeout{}
  \typeout{******************************************}
  \typeout{** Please run "bibtex \jobname" to optain}
  \typeout{** the bibliography and then re-run LaTeX}
  \typeout{** twice to fix the references!}
  \typeout{******************************************}
  \typeout{}
 }

\end{document}